\documentclass[floatfix,twocolumn,showpacs,preprintnumbers,amsmath,amssymb,aps,pra,superscriptaddress,longbibliography]{revtex4-1}

\usepackage[version=3]{mhchem}
\usepackage{chemformula}
\usepackage{graphicx}% Include figure files
\usepackage{dcolumn}% Align table columns on decimal point 
\usepackage{bm}% bold math
\usepackage{float} 
\usepackage{lipsum}
\usepackage{enumitem,amssymb} 
\usepackage{color}
\usepackage{siunitx}

\usepackage{pifont}

\usepackage{leftidx}
\usepackage{tabulary}
\usepackage{tabularx,booktabs}

\usepackage{xparse}
\usepackage{siunitx} 

\usepackage[utf8]{inputenc}
\usepackage[T1]{fontenc}
\usepackage{csquotes} 
\DeclareUnicodeCharacter{2009}{\,}

\newcommand{\br}{\bm{r}}

\newcommand{\s}{_\mathrm{{\scriptscriptstyle S}}}
\newcommand{\h}{_\mathrm{{\scriptscriptstyle H}}}
\newcommand{\xc}{_\mathrm{{\scriptscriptstyle XC}}}

\newcommand{\PBE}{^\mathrm{{PBE}}}
\newcommand{\tPBE}{^\mathrm{{tPBE}}}
\newcommand{\unif}{\mathrm{{unif}}}

\begin{document}

\title{Electrical Conductivity of Warm Dense Hydrogen from Ohm's Law and Time-Dependent Density Functional Theory}

\author{Kushal Ramakrishna}
\email{k.ramakrishna@hzdr.de} 
\affiliation{Center for Advanced Systems Understanding (CASUS), D-02826 G\"orlitz, Germany}
\affiliation{Helmholtz-Zentrum Dresden-Rossendorf (HZDR), D-01328 Dresden, Germany}

\author{Mani Lokamani}
\affiliation{Helmholtz-Zentrum Dresden-Rossendorf (HZDR), D-01328 Dresden, Germany}

\author{Attila Cangi} 
\email{a.cangi@hzdr.de}
\affiliation{Center for Advanced Systems Understanding (CASUS), D-02826 G\"orlitz, Germany}
\affiliation{Helmholtz-Zentrum Dresden-Rossendorf (HZDR), D-01328 Dresden, Germany}

\date{\today}% It is always \today, today, %  but any date may be explicitly specified
             
\begin{abstract} 
Understanding the electrical conductivity of warm dense hydrogen is critical for both fundamental physics and applications in planetary science and inertial confinement fusion. We demonstrate how to calculate the electrical conductivity using the continuum form of Ohm's law, with the current density obtained from real-time time-dependent density functional theory. This approach simulates the dynamic response of hydrogen under warm dense matter conditions, with temperatures around 30,000 K and mass densities ranging from 0.02 to 0.98 g/cm³. We systematically address finite-size errors in real-time time-dependent density functional theory, demonstrating that our calculations are both numerically feasible and reliable. Our results show good agreement with other approaches, highlighting the effectiveness of this method for modeling electronic transport properties from ambient to extreme conditions.
\end{abstract}    
 
\maketitle

%%%%%%%%%%%%%%%%%%%%%%%%%%%%%%%%%%%%%%%%%%%%%%%%%%%%%%%%%%%%%%%%%%%%%%%%%%%%%%%
\section{Introduction}   
%%%%%%%%%%%%%%%%%%%%%%%%%%%%%%%%%%%%%%%%%%%%%%%%%%%%%%%%%%%%%%%%%%%%%%

Warm dense hydrogen refers to a state of hydrogen where it exists at high temperatures and pressures, typically in the range of a few thousand Kelvin and several gigapascals of pressure~\cite{graziani2014frontiers,bonitz2020ab,bonitz2024principlessimulationsdensehydrogen}. In this state, hydrogen exhibits properties that are intermediate between those of dense plasma and condensed matter. This is of interest for understanding the interiors of gas giant planets like Jupiter and Saturn~\cite{militzer1,saumon1}, as well as in the core of brown dwarfs~\cite{booth2015laboratory}, and for technological applications such as inertial confinement fusion~\cite{nuckolls1972laser,betti2016inertial}, where a fuel capsule passes through the warm dense matter phase before ignition. This process is achieved in the laboratory using high-energy laser pulses or by compressing hydrogen with powerful static compression techniques~\cite{PhysRevE.90.013104,zastrau_resolving_2014,Fletcher2015,falk2018experimental}. Understanding the transport and optical properties of warm dense matter is therefore essential for accurate modeling of the inertial confinement fusion process, particularly for modeling the radiation transport processes of an imploding fuel capsule. 
Experimental investigations have been conducted to study the electrical conductivity of warm dense matter with techniques such as laser-driven shock compression, pulsed power devices, static compression~\cite{PhysRevLett.41.994,PhysRevLett.76.1860,PhysRevB.59.3434,nellis1992electronic} and recently using terahertz transmission measurements~\cite{chen2021ultrafast,ofori2024dc}, but these are challenging. First-principles modeling techniques are therefore of great use in completing the picture provided by the experimental data~\cite{edrwdm}.

The Kubo-Greenwood formula~\cite{doi:10.1143/JPSJ.12.570,Greenwood_1958}, derived in the framework of linear response theory, is the standard method for calculating electrical conductivity. It has been used extensively in Kohn-Sham density functional theory (DFT) to calculate the frequency-dependent optical conductivity of high-pressure hydrogen, particularly in its liquid state~\cite{PhysRevE.54.2844,pfaffenzeller1997structure,PhysRevB.63.184110,PhysRevB.82.195107,PhysRevB.77.184201,PhysRevB.83.235120,PhysRevE.105.065204,lambert2011transport}. The Kubo-Greenwood formula allows the calculation of the optical conductivity by integration over the energy spectrum, taking into account the differences in the Fermi-Dirac distribution between different Kohn-Sham orbitals~\cite{PhysRev.140.A1133}.

More recently, the stochastic DFT approach has also been used to calculate the conductivity using the Kubo-Greenwood method~\cite{PhysRevLett.111.106402,PhysRevB.97.115207,PhysRevE.109.065304}. Although quantum Monte Carlo methods have been applied to compute the electrical conductivity, they remain computationally expensive. Quantum Monte Carlo faces significant challenges in accurately resolving excited states due to the fermion sign problem and finite-size effects~\cite{morales2010evidence,PhysRevLett.103.256401}.

Alternatively, the electrical conductivity can be calculated directly using the continuum form of Ohm's law~\cite{PhysRevB.107.115131}. This is achieved by the real-time formalism of time-dependent density functional theory (TDDFT)~\cite{PhysRevB.54.4484}. The response function obtained naturally incorporates electron-electron correlation effects~\cite{baczewski2021predictions,PhysRevLett.116.115004,PhysRevB.107.115131,Ramakrishna2023iop}, which are absent in the conventional approaches where the Kubo-Greenwood formula is applied directly to Kohn-Sham orbitals.

Real-time TDDFT has previously been used to model collective effects and conductivity in iron~\cite{PhysRevB.107.115131,Ramakrishna2023iop,PhysRevMaterials.8.033803,nikolov2023probingironearthscore}, although its application has been limited to relatively small system sizes and modest external conditions due to the high computational cost. 

In this work, we demonstrate the ability to evaluate the electrical conductivity of warm dense hydrogen using TDDFT, explicitly addressing finite-size errors. This step brings the method to the same practical level as the widely used Kubo-Greenwood approach, while providing a more accurate treatment of electron-electron interactions through the inclusion of the exchange-correlation (XC) kernel. Furthermore, we study the impact of temperature on the conductivity in the context of thermal XC effects, which are particularly important for warm dense hydrogen~\cite{PhysRevB.101.195129}.
 
%%%%%%%%%%%%%%%%%%%%%%%%%%%%%%%%%%%%%%%%%%%%%%%%%%%%%%%%%%%%%%%%%%%%%%%%%%%%%%
%\section{Methods}
\section{Methodological and Computational Details}
%%%%%%%%%%%%%%%%%%%%%%%%%%%%%%%%%%%%%%%%%%%%%%%%%%%%%%%%%%%%%%%%%%%%%%%%%%%%%%

\begin{figure*}[t]  
\centering       
\includegraphics[width=0.85\linewidth]{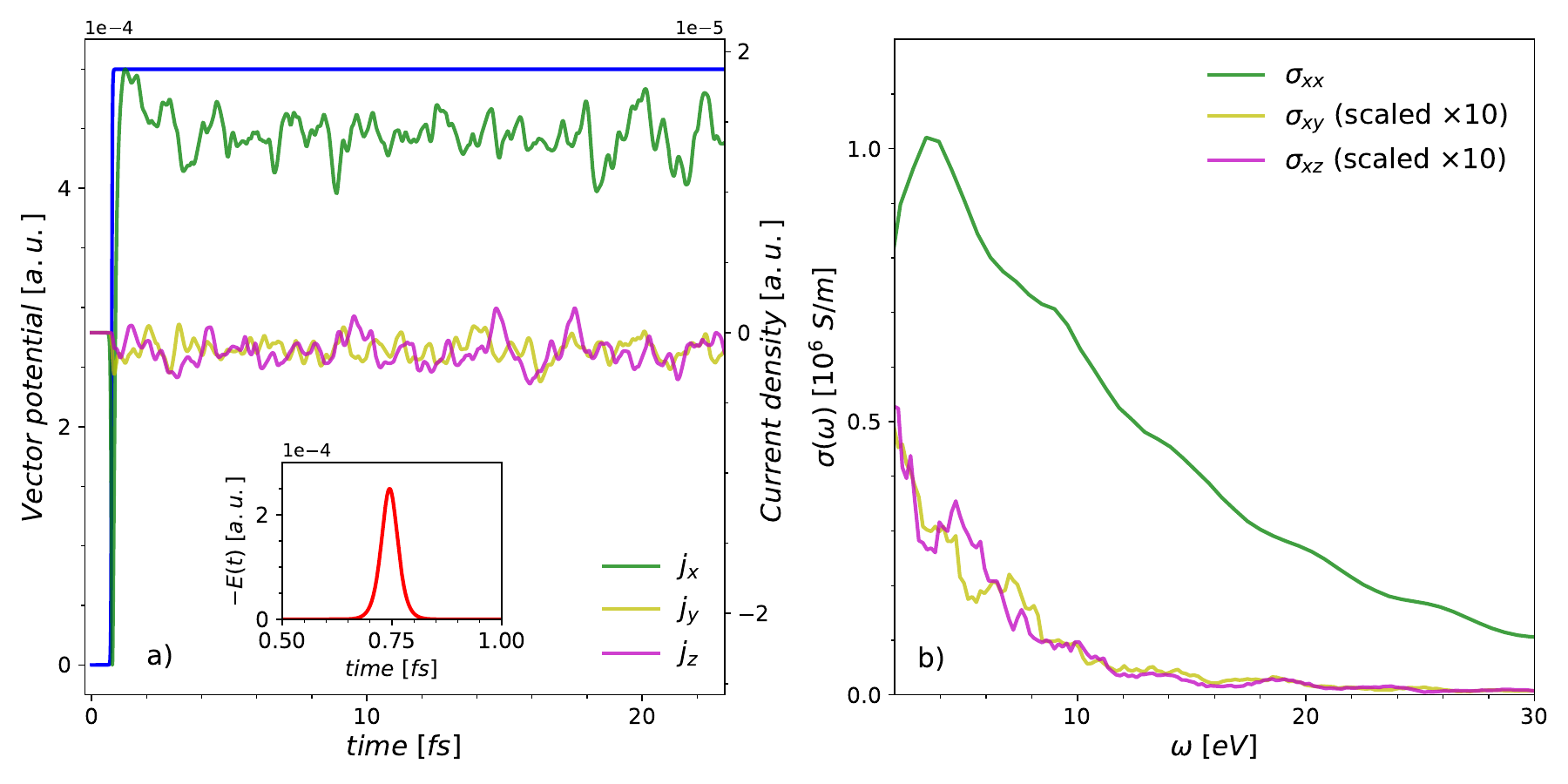} 
\caption{Calculating the electrical conductivity using the continuum form of Ohm's law: (a) applied vector potential (blue) along the $x$ direction as a function of time, along with the induced current density (green) in the $x$ direction. The yellow and purple curves represent the induced current densities along the $y$ and $z$ directions, respectively, both of which are negligible compared to the $x$ component. The inset provides a closer view of the corresponding electric field (red) near the beginning of the time propagation where the delta-kick pulse is applied; (b) dynamic electric conductivity component $\sigma_{xx}$ (green) obtained using Eq.~\ref{eq.ohmslaw.ft}. The off-diagonal components $\sigma_{xy}$ and $\sigma_{xz}$ (yellow and purple) are also shown, scaled by a factor of ten for clarity.}
\label{rt_theory}  
\end{figure*}

We utilize real-time TDDFT to compute the electrical conductivity based on the continuum form of Ohm's law:

\begin{equation} 
j_{i}(\br, t) = \sum_{j} \int d\br' \int_{-\infty}^{t} dt'\ \sigma_{ij}(\br,\br',t-t')\, E_{j}(\br',t')\ . 
\label{eq.ohmslaw} 
\end{equation}

In this expression, an external electric field $\boldsymbol{E}(\br', t) = -\partial \boldsymbol{A}(\br', t)/\partial t$ induces a current density $\boldsymbol{j}(\br, t)$, where $\boldsymbol{A}(\br', t)$ denotes the corresponding external vector potential. To achieve this, we solve the time-dependent Kohn-Sham equations:

\begin{equation} 
\hat{H}\s \psi_{n,k}(\br,t) = i \frac{\partial}{\partial t} \psi_{n,k}(\br,t) 
\label{eq} 
\end{equation}

for the Kohn-Sham orbitals $\psi_{n,k}(\br,t)$, with the effective Hamiltonian:

\begin{equation} 
\hat{H}\s = \frac{1}{2} \left[-i \nabla + \boldsymbol{A}\s(\br,t)\right]^{2} + v\s(\br,t)\ .
\label{eq2} 
\end{equation}

Here, $v_{s}(\br,t) = v_{ext}(\br,t) + v\h(\br,t) + v\xc(\br,t)$ is the Kohn-Sham potential, which combines the external, Hartree, and XC potentials. The effective vector potential $\boldsymbol{A}\s(\br,t) = \boldsymbol{A}(\br,t) + \boldsymbol{A}\xc(\br,t)$ includes both the external and XC contributions. Solving these equations yields the time-dependent Kohn-Sham orbitals from which the induced current density is calculated:

\begin{equation} 
\boldsymbol{j}(\br,t) = \Im \left[\sum_{i}^{N} \psi_{n,k}^{*}(\br,t) \nabla \psi_{n,k}(\br,t)\right] + n(\br,t) \boldsymbol{A}\s(\br,t)\ .
\label{curr_dens} 
\end{equation}

Assuming a spatially constant electric field and treating its time-dependence within the dipole approximation, the dynamical electrical conductivity is obtained as:

\begin{equation} 
\sigma_{ij}(\omega) = \frac{\tilde{j}_{i}(\omega)}{\tilde{E}_{j}(\omega)}\ , 
\label{eq.ohmslaw.ft}
\end{equation}

where we integrated Eq.~(\ref{eq.ohmslaw}) over the spatial coordinates and performed Fourier transforms over time. The current and electric field are given by $\tilde{j}_{i}(\omega) = \int dt \int d\br\ j_{i}(\br,t) e^{i \omega t}$ and $\tilde{E}_{j}(\omega) = \int dt\ E_{j}(t) e^{i \omega t}$, respectively. Note that the current and electric field are vectors, while the electrical conductivity is a tensor. Throughout this work, we use atomic units (a.u.) for the real-time TDDFT simulations.
The total frequency-dependent conductivity is obtained from Eq.~(\ref{cond_avg}) using the trace of the conductivity tensor:

\begin{equation} 
\sigma(\omega) = \frac{1}{3\mathnormal{N}} \sum\limits_{a=1, ~ i=x,y,z }^{a=\mathnormal{N}} \sigma_{ii}^{a}(\omega)\ .
\label{cond_avg} 
\end{equation}

The key macroscopic quantity of interest is the electrical conductivity at zero frequency, known as the DC conductivity, $\sigma_{dc}$. We determine it by fitting the frequency-dependent conductivity, $\sigma(\omega)$, to a Drude model: $\sigma(\omega) = \sigma_{dc}/(1 + \omega^{2} \tau^{2})$, with particular focus on the low-energy region, where, $\tau$ denotes the relaxation time~\cite{ashcroft2022solid}.

The process of calculating the electrical conductivity using the continuum form of Ohm's law is illustrated in Fig.~\ref{rt_theory}. Running a real-time TDDFT simulation begins with the calculation of the initial state, which is typically a set of Kohn-Sham orbitals obtained from ground-state or thermal DFT calculations. These are performed with an in-house modified version of the FHI-AIMS code which uses a numerical atom-centered basis function framework~\cite{blum2009ab,hekele2021all}. The modifications to this code allow the use of the thermal PBE functional (see Appendix), making it possible to quantify thermal XC effects on the electrical conductivity.

In our real-time TDDFT simulations, we typically apply a delta-kick pulse with a weak amplitude $\boldsymbol{E}_{0}$ of the form:

\begin{equation}
\boldsymbol{E}(t) = \boldsymbol{E}_{0} \left( e^{ \frac{t-t_{0}}{t_{w}}}  \right)
\left(1 +  e^{\frac{t-t_{0}}{t_{w}}}\right)^{-2}\ .
\end{equation}

For the example shown in Fig.~\ref{rt_theory}, we apply a vector potential along the $x$ axis. The inset shows the corresponding electric field (blue curve) applied as a delta-kick pulse. 

The pulse is centered at $t_{0} = 30$ a.u. with a width of $t_{w} = 0.5$ a.u. (1 a.u. $= 24.819$ attoseconds) and an amplitude of $\boldsymbol{E}_{0} = 0.001$ a.u. (1 a.u. $= 5.142 \times 10^{11}$ V/m) within the
weak perturbation limit. The total simulation time is 110 fs, which gives an energy resolution of about 0.04 eV. Sufficiently long real-time propagation is critical to ensure that the induced electric current reaches a stable state and that adequate energy resolution is achieved. The Crank-Nicholson scheme~\cite{crank1947practical} is used for time evolution, allowing a time step $\Delta t$ of up to 0.1 a.u. without sacrificing accuracy. 
The Perdew-Burke-Ernzerhof (PBE) functional~\cite{PBE96} is used to approximate the XC potential in Eq.~(\ref{eq}) within the adiabatic approximation.  

The induced current density, calculated from the time evolution of the Kohn-Sham orbitals using Eq.~(\ref{curr_dens}), is plotted for all three spatial dimensions. As expected, the primary current is induced along the $x$ direction (green curve), while the currents in the $y$ and $z$ directions are negligible (yellow and purple curves).

The resulting component of the electrical conductivity tensor, $\sigma_{xx}$ (green), derived from Eq.~(\ref{eq.ohmslaw.ft}), is shown in Fig.~\ref{rt_theory}b. In addition, the off-diagonal components, $\sigma_{xy}$ and $\sigma_{xz}$ (yellow and purple curves), are shown to be negligible compared to $\sigma_{xx}$, as expected.

The calculation of electrical conductivity typically assumes a fixed ionic configuration. However, in disordered systems such as warm dense matter, the positions of the ions have a substantial effect on the resulting conductivity. To account for this disorder, we sample 8 to 10 different ionic configurations for a given mass density and temperature. Real-time TDDFT calculations are performed on each configuration separately, and the electrical conductivity is obtained by averaging over the sampled configurations.

The ionic configurations used to sample conductivity in Eq.~(\ref{cond_avg}) are generated from density functional molecular dynamics (DFT-MD) simulations, performed using the VASP~\cite{KH93,KF96,KF96b,KJ99} code. These simulations utilize PAW pseudopotentials (PAW\_PBE\_H\_h\_GW and PAW\_PBE)~\cite{B94} with the PBE approximation~\cite{PBE96}. A plane wave cutoff of 1200~eV is applied, and the convergence criterion for each self-consistency cycle is set to $10^{-5}$.
In these calculations, we employ the Mermin formulation of thermal DFT, incorporating Fermi-Dirac occupations for the electronic states~\cite{M65}. The simulations are conducted in the NVE ensemble, controlled by a Nose-Hoover thermostat~\cite{H85}. Depending on the density, the ionic time step $\Delta t$ ranges from 0.04 to 0.14~fs, with total simulation times extending up to 20,000 steps. The ionic configurations used for conductivity calculations are sampled from the final few thousand steps of the DFT-MD simulations.

%%%%%%%%%%%%%%%%%%%%%%%%%%%%%%%%%%%%%%%%%%%%%%%%%%%%%%%%%%%%%%%%%%%%%%%%%%%%%%
\section{Results}

\begin{figure}[t]
\includegraphics[width=1.0\columnwidth]{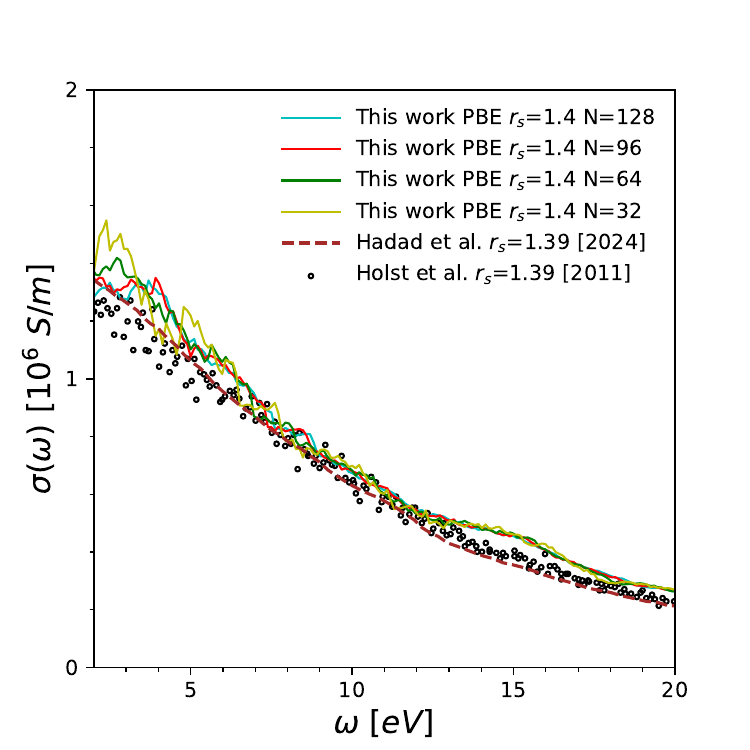}
\caption{Assessing finite-size effects in the frequency-dependent conductivity of hydrogen at a density of $r\s$=1.4 and a temperature of T=30,000~K, by increasing the number of atoms in the simulation from 32 to 128. The PBE functional was used in our real-time TDDFT calculations. Our results are compared with the Kubo-Greenwood method from Refs.~\onlinecite{PhysRevB.83.235120,PhysRevE.109.065304}, which used DFT-based orbitals with PBE and LDA functional respectively. Note that in the Kubo-Greenwood calculations, the density was slightly different, with $r\s$=1.39.}   
\label{rs1p4_sigma}
\end{figure} 

\begin{figure}[t]
\includegraphics[width=1.0\columnwidth]{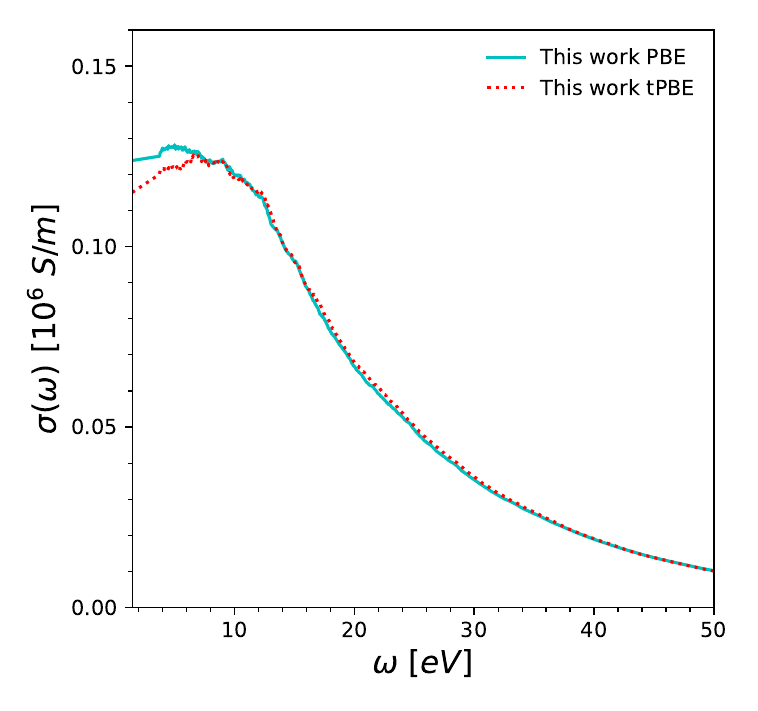}
\caption{Assessing thermal XC effects in the frequency-dependent conductivity of hydrogen which is evaluated using real-time TDDFT using PBE and thermal PBE (tPBE) functionals at $r\s$=2.2, T=30,000~K.}  
\label{thermalxc}
\end{figure} 

We begin by analyzing finite-size effects in the evaluation of frequency-dependent conductivity using real-time TDDFT. These calculations are computationally intensive, which has limited the number of atoms in previous studies to a few tens~\cite{PhysRevB.107.115131,Ramakrishna2023iop}. This is significantly smaller than the system sizes typically used in the conventional Kubo-Greenwood approach, where a few hundred atoms are commonly considered. However, the efficient implementation of real-time TDDFT in FHI-AIMS allows us to scale up the number of atoms to those used in Kubo-Greenwood calculations.

For this study, we investigate finite-size effects by considering hydrogen at a density of $r\s$=1.4 ($\rho$=0.98~g/cm$^{3}$) and a temperature of T=30,000~K. We examine system sizes ranging from 32 to 128 atoms, with results shown in Fig.~\ref{rs1p4_sigma}. All real-time TDDFT calculations are performed with the PBE functional. Additionally, the black dots in the figure represent conductivity values using the Kubo-Greenwood method on Kohn-Sham orbitals~\cite{PhysRevB.83.235120} for comparison at a slightly higher density ($\rho$=1.0~g/cm$^{3}$, $r\s$=1.39). The brown curve shows the conductivity based on the Kubo-Greenwood formula using stochastic DFT~\cite{PhysRevE.109.065304}.

Our real-time TDDFT results show good agreement with Kubo-Greenwood calculations, both with Kohn-Sham orbitals and stochastic DFT, for similar system sizes. For the smallest system size (N=32), the DC conductivity is higher and shows noticeable fluctuations. However, from N=64, the DC conductivity converges well, with larger system sizes leading to smaller fluctuations. Although our results are consistent with previous studies, it is important to note that under the density conditions studied ($\rho$=0.98~g/cm$^{3}$), real-time TDDFT simulations using a numerical atom-centered basis set are more computationally demanding than the Kubo-Greenwood method, which uses a plane-wave basis in a density functional theory code such as VASP.
Despite the higher computational cost, real-time TDDFT simulations converge faster in terms of propagation time when using N=128 atoms compared to smaller system sizes such as N=32. 

These results demonstrate that real-time TDDFT is a scalable and effective method for calculating the electrical conductivity by tracking the time evolution of the current density, placing it alongside the Kubo-Greenwood method as a state-of-the-art approach. It is important to note, however, that the results of real-time TDDFT are not identical to those of the Kubo-Greenwood formula. Due to the inclusion of the XC kernel, real-time TDDFT provides a more accurate representation of the electron-electron interactions than the Kubo-Greenwood approach when evaluated on Kohn-Sham orbitals, as is commonly done. This difference becomes particularly significant in systems where electron-electron interactions play a larger role, such as at lower densities~\cite{Gross2012}. Additionally, it offers advantage over Kubo-Greenwood method as it is not limited to linear-response regimes and does not rely on the calculation of unoccupied states.

As the system temperature increases, thermal XC effects become increasingly important. XC functionals that incorporate an explicit temperature dependence~\cite{PhysRevLett.112.076403,PhysRevLett.120.076401,PhysRevLett.119.135001,PhysRevB.101.245141} have been shown to play a crucial role in the accurate evaluation of the equation of state~\cite{PhysRevE.93.063207,PhysRevLett.120.076401,PhysRevB.101.195129,bonitz2024principlessimulationsdensehydrogen} and Hugoniot curves~\cite{PhysRevB.99.214110,PhysRevB.107.155116,bonitz2024principlessimulationsdensehydrogen}.
More recently, the influence of thermal XC effects on optical properties has also been studied, particularly within the framework of the Kubo-Greenwood formula~\cite{PhysRevE.93.063207,PhysRevB.105.L081109,PhysRevB.107.155116}. 

In the following, we investigate the impact of thermal XC effects on both the frequency-dependent and DC conductivities. To this end, we utilize a recent development: the thermal PBE (tPBE) functional, the thermal analog of the well-known PBE functional, derived from conditional probability density functional theory~\cite{kozlowski2023generalized}. Since the derivation, implementation, and benchmarking of the tPBE functional for static properties is detailed elsewhere~\cite{kozlowski2023generalized,RaLoMo2024}, we provide only brief information about the functional in the Appendix.

As the system approaches the Fermi temperature ($T_F$), thermal XC effects become increasingly important. To explore their influence on the electrical conductivity, we study hydrogen at an isotherm of T=30,000 K (0.25 $T_{F}$) and a density of $r\s$=2.2 ($\rho = 0.25$ g/cm$^3$) using a system size of 128 atoms. The frequency-dependent conductivity is calculated using both the PBE and tPBE functionals and is shown in Fig.~\ref{thermalxc}. The results from both functionals agree well over most of the frequency range, with notable deviations at lower frequencies. These differences are due to a shift in the Kohn-Sham eigenvalues in the tPBE calculations compared to the PBE calculations.

\begin{figure}
\includegraphics[width=1.0\columnwidth]{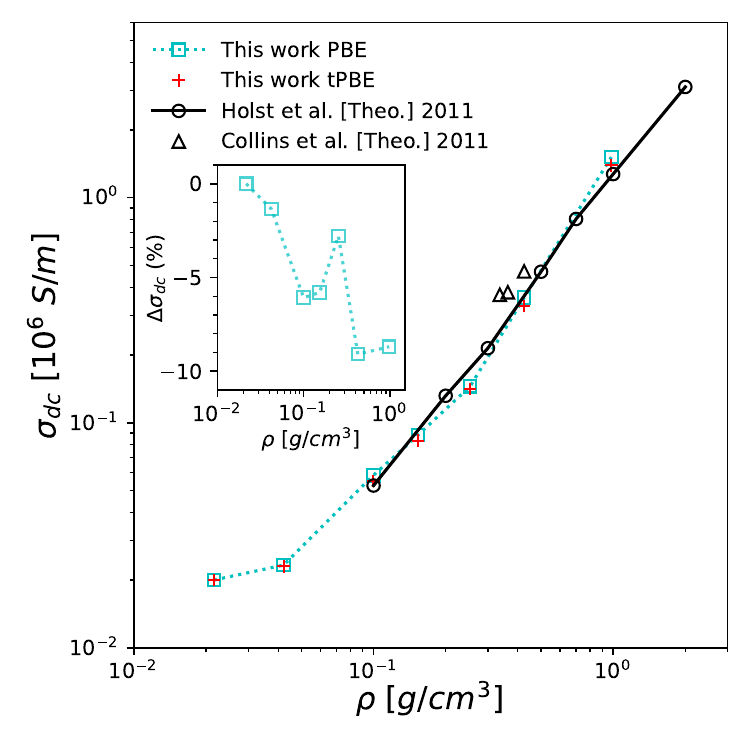}
\caption{DC conductivity of hydrogen as a function of density at an isotherm T=30,000~K using PBE and tPBE. Theoretical data stems from Refs.~\cite{PhysRevB.83.235120,PhysRevB.63.184110} evaluated with Kubo-Greenwood method using PBE functional. The inset plot shows the relative difference in conductivity between tPBE and PBE from this work.}
\label{isotherm_30000k}
\end{figure}

In Fig.~\ref{isotherm_30000k} we show the DC conductivity of hydrogen along a range of densities at the same isotherm, T=30,000 K. As before, we use a 128-atom simulation cell, with further details on computational performance provided in the Appendix. Both the PBE (blue squares) and tPBE (red crosses) results show good agreement with previous Kubo-Greenwood results from Holst \textit{et al.}~\cite{PhysRevB.83.235120} (black curve) and Collins \textit{et al.}~\cite{PhysRevB.63.184110} (black triangles). The inset compares the PBE and tPBE results, showing at most a 9\% difference. In particular, the tPBE calculations consistently yield a lower DC conductivity, in better agreement with the Kubo-Greenwood data of Holst \textit{et al.}~\cite{PhysRevB.83.235120}. This observation is consistent with previous results on aluminum using thermal functionals~\cite{PhysRevE.93.063207,PhysRevB.105.L081109}.

\section{Conclusions} 

In this work, we have demonstrated the capability of real-time TDDFT to calculate the electrical conductivity of warm dense hydrogen from the continuum form of Ohm's law, with particular emphasis on finite-size effects and thermal XC contributions. 
The use of numerical atom-centered basis functions, as implemented in FHI-AIMS, allows for computationally demanding calculations and scaling of the required TDDFT calculations to system sizes comparable to those used in the conventional Kubo-Greenwood approach. Our results show that real-time TDDFT provides an accurate and scalable method for evaluating electrical conductivity, yielding results that are in good agreement with the Kubo-Greenwood method. However, real-time TDDFT has the advantage of more accurately accounting for electron-electron interactions through the inclusion of the XC kernel.

We also investigated the role of thermal XC effects in both frequency-dependent and DC conductivities. Using the recently developed thermal PBE functional, we investigated the impact of thermal XC effects on the conductivity, particularly near the Fermi temperature. Our results show that the tPBE functional shifts the Kohn-Sham eigenvalues, particularly at lower frequencies, leading to a reduction in the DC conductivity compared to the standard PBE functional.

Overall, our work reinforces the growing utility of real-time TDDFT for studying the transport properties of warm dense matter, providing a more accurate treatment of electron-electron interactions and thermal effects. These results are essential for advancing the understanding of transport properties under extreme conditions, with implications for fields ranging from planetary science to inertial confinement fusion. Future work can extend these investigations to other materials and conditions, and explore the computational efficiency and scalability of real-time TDDFT for larger systems, as well as different temperatures and densities.

\begin{acknowledgments} 

We are grateful to J. K. Dewhurst for the hospitality at Max-Planck-Institut f\"{u}r Mikrostrukturphysik Halle and for stimulating discussions.

This work was partially supported by the Center for Advanced Systems Understanding (CASUS) which is financed by Germany's Federal Ministry of Education and Research (BMBF) and by the Saxon state government out of the State budget approved by the Saxon State Parliament. Computations were performed on a Bull Cluster at the Center for Information Services and High-Performance Computing (ZIH) at Technische Universit\"at Dresden and on the cluster Hemera at Helmholtz-Zentrum Dresden-Rossendorf (HZDR). 
\end{acknowledgments}    

\section{Appendix}
\label{appendix}

\subsection{Computational Performance}

\begin{figure}
\includegraphics[width=1.0\columnwidth]{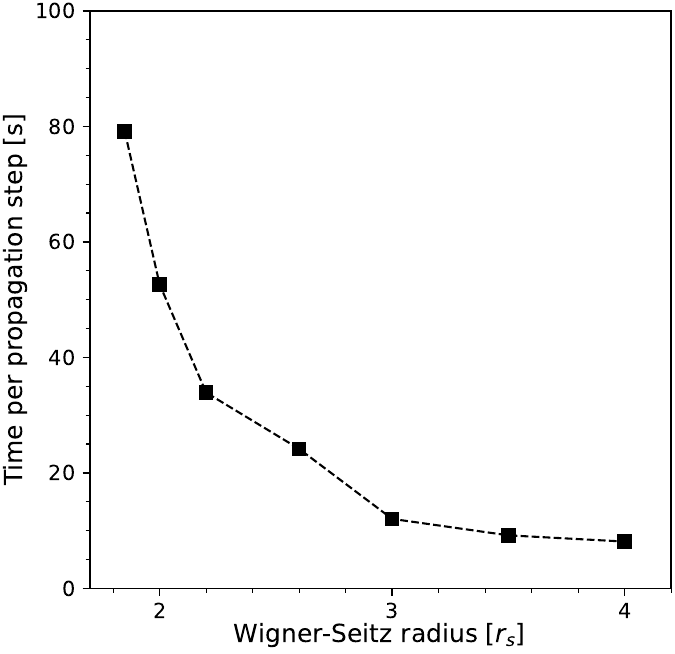}
\caption{Mean time in seconds required per real-time propagation step based on densities ($r\s$) using PBE XC functional.}
\label{scaling}
\end{figure}

TDDFT simulations were carried out using AMD Epyc Genoa 9654~@2.4~GHz (96 cores per node). The performance metrics for our simulations are shown in Fig.~\ref{scaling} where we compare the mean time required per real-time propagation step for a range of densities ($r\s$) using PBE XC functional. Propagation timestep  $\Delta t$=0.1~a.u. (1 a.u.=24.819 attoseconds) is considered throughout. Similar numbers of CPUs and system sizes (N=128) are taken into account in the simulations for uniformity and performance scaling. Note that, for lower densities, the numerical atom-centered basis functions are advantageous over traditional planewave methods.

\subsection{Comparison Across DFT Software Packages}

\begin{figure}
\includegraphics[width=1.0\columnwidth]%
{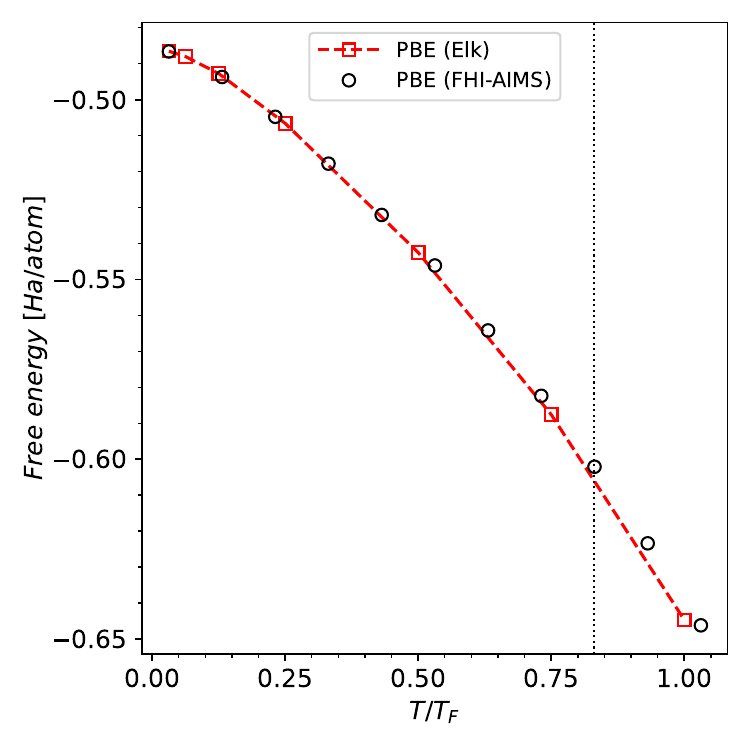}
\caption{Comparison of the free energy of hydrogen at $r\s$=4 as a function of reduced temperature ($\theta=T/T_{F}$) using PBE functional evaluated with Elk and FHI-AIMS. The vertical line indicates the conditions of reduced temperature $T/T_{F}~\sim~0.83$ corresponding to T=30,000~K at $r\s$=4 considered in this work.}  
\label{rs4_elk_fhiaims}
\end{figure}

Figure~\ref{rs4_elk_fhiaims} shows the free energy of hydrogen at $r\s$=4 ($\rho$=0.042~g/cm$^{3}$) as a function of reduced temperature $\theta=T/T_{F}$ where $T_{F}$ is the Fermi temperature, and, $T$ is the temperature.  The simulations are performed using FHI-AIMS within a numeric atom-centered basis function framework~\cite{blum2009ab,hekele2021all} and full potential linearized augmented-plane-wave (FP-LAPW) method~\cite{singh2006planewaves} implemented in the highly accurate Elk code~\cite{elk}. We obtain good agreement in the evaluation of free energy using PBE across the DFT packages.

\subsection{Finite-Temperature Exchange-Correlation Effects}
\label{app.fin.temp.xc}
Additionally, to consider thermal XC effects, we evaluate the conductivity with generalization of the PBE functional to finite-temperature based on the conditional-probability density functional methodology~\cite{kozlowski2023generalized}. 
Accordingly, the functional form of the thermal PBE enhancement factor~\cite{kozlowski2023generalized} is given by

\begin{equation}
f\xc^{\tPBE}(r\s,\nu, S, \theta) = \frac{f\xc^{\unif}(r\s, \nu, \theta)}{f\xc^{\unif}(r\s, \nu)}\,f\xc^{\PBE}(r\s,\nu,S)\,,
\label{enhancement_factor_eqn}
\end{equation}

where $f\xc^{\unif}(r\s, \nu, \theta)$ denotes the enhancement factor of the uniform electron gas at finite--temperature~\cite{PhysRevLett.119.135001} and $f\xc^{\unif}(r\s, \nu)$ in the ground state~\cite{perdew_wang}. $r\s=(3/4 \pi n)^{1/3}$ is the Wigner-Seitz radius, $\nu=(n_{\uparrow} -{n_\downarrow})/n$ the relative spin orientation defined in terms of spin-up and spin-down densities, $\theta = T/T_{F}$ the reduced temperature with $T$ the electronic temperature and $T_{F}$ the Fermi temperature, $S=|\nabla n|/(2k_{F} n)$ is the dimensionless density gradient with $k_{F}$ the Fermi wavevector. The thermal prefactor takes into account the temperature dependence of the thermal PBE functional. As a result, the consequent XC free energy of thermal PBE drops to PBE at zero temperature.
We have implemented the thermal PBE functional in an in-house version of LIBXC (library of exchange-correlation functionals)~\cite{lehtola2018recent,marques20122272}. 
Further details of the implementation and benchmarks of static properties using thermal DFT will be provided in a forthcoming work~\cite{RaLoMo2024}.

%%%%%%%%%%%%%%%%%%%%%%%%%%%%%%%%%%%%%%%%%%%%%%%%%%%%%%%%%%%% 
\clearpage

\bibliography{bibliography}

\end{document}